\documentclass[twocolumn,aps,prl]{revtex4} 
\usepackage{graphics,amsmath,graphicx} 

\begin{document} 
\title{Polaronic signatures in the optical properties 
of Nd$_{2-x}$Ce$_x$CuO$_4$} 
\author{E. Cappelluti$^{1,2}$, S. Ciuchi$^{3,1}$, and S. Fratini$^{4}$} 
\thanks{On leave of absence at the ICMM-CSIC, Madrid, Spain.}

\affiliation{$^1$SMC Research Center and ISC, INFM-CNR, v. dei Taurini 19, 
00185 Rome, Italy}
 
\affiliation{$^2$Dipartimento di Fisica, Universit\`a ``La Sapienza'', 
P.le A. Moro 2, 00185 Rome, Italy} 

\affiliation{$^3$CNISM and 
Dipartimento di Fisica, Universit\`a dell'Aquila, 
via Vetoio, I-67010 Coppito-L'Aquila, Italy}

\affiliation{$^4$Institut N\'eel - CNRS \& Universit\'e Joseph Fourier,
BP 166, F-38042 Grenoble Cedex 9, France}

\begin{abstract} 
We investigate the  temperature and doping dependence of the optical
conductivity $\sigma(\omega)$ of Nd$_{2-x}$Ce$_x$CuO$_4$ in terms of
magnetic/lattice polaron formation. We employ dynamical mean-field theory
in the context of the Holstein-$t$-$J$ model
where an exact analytical solution is
available in the limit of infinite connectivity.  We show that the pseudogap
features in the optical conductivity of this compound can be associated to the
formation of lattice polarons assisted by the magnetic interaction.
\end{abstract} 
%\pacs{78.20.Bh, 71.38.-k, 71.10.Fd}
\date{\today} 
\maketitle 

Understanding the phase diagram of high-$T_c$ superconductors
still represents, twenty years after their discovery,
one of the main challenges in this field.
One of the reasons of such difficulties relies on the fact that 
different interactions are simultaneously operating,  involving
different degrees of freedom.  On one hand, evidences are gathering,
from different experimental techniques, about
a relevant role of the electron-lattice
interaction \cite{lanzara,gweon,khasanov1,khasanov2,davis}.
On the other hand, short-range antiferromagnetic (AF) correlations,
ruled by the exchange coupling $J$,
are still present in the normal state even when true long-range
AF ordering is destroyed.
The interplay between these two kinds of interactions has been recently
pointed out in undoped and weakly doped cuprates,
where angle-resolved photoemission
spectroscopy (ARPES) revealed a broad high-energy peak
well distinguished from the quasi-particle states with vanishingly small
spectral weight \cite{shen,shen2}.
The broadness of such peak has been interpreted in terms
of a multi-phonon Franck-Condon
structure typical of a lattice polaron, superposed on the dynamics
of a single hole in the $t$-$J$
model \cite{shen,shen2,mish1,mish2,cataudella}.
The validity of this scenario at finite doping, and its relation
with the pseudogap observed by means of different techniques
is however not clear. 

Among the different families of copper oxides, electron-doped cuprates 
represent the most suitable systems where to investigate the
interplay between magnetic and lattice polarons.
This is because, unlike hole-doped systems,
the long-range AF ordering extends up to quite large doping levels
$\delta \approx 0.14$, so that the relative importance of 
long-range {\it vs.} short-range magnetic correlations can be more
clearly addressed.
The optical conductivity (OC) of the electron-doped compound
Nd$_{2-x}$Ce$_x$CuO$_4$  has 
been systematically  analyzed 
as function of temperature and doping 
in Ref. \onlinecite{onose}. There,
a mid-infrared (MIR) shoulder associated with a pseudogap
state has been observed in the spectra at finite doping.
Quite interestingly, they observed: $i$) a shift of the
MIR peak towards lower frequencies upon increasing $\delta$, 
corresponding to a pseudogap closing
with doping; $ii$) a filling of the pseudogap upon increasing the temperature;
$iii$) the absence of any signature
of the N\'eel temperature in the temperature evolution of the 
pseudogap spectral weight (defined as the spectral weight in
the region between the MIR shoulder and a low-energy infrared
cut-off $\omega_{\rm min}\simeq 0.12$ eV), whereas
a remarkable
non-monotonic kink is present at a higher temperature $T^*$.
The latter feature suggests 
a prominent role of short-range correlations 
in governing the observed spectral properties.

A detailed theoretical study of the polaron features in the optical
conductivity of one hole in the Holstein-$t$-$J$ model,
using dynamical mean-field theory (DMFT),
has been recently reported in Ref. \cite{ccf}.
In this Letter,  we apply such approach to analyze
the optical data of electron-doped cuprates Nd$_{2-x}$Ce$_x$CuO$_4$ (NCCO),
where exhaustive measurements as function of both temperature and doping
are available \cite{onose}. We show that the optical conductivity
in these compounds can be naturally interpreted in terms
of spin/lattice polarons resulting from the joint positive cooperation
between the electron-phonon (el-ph) and magnetic 
exchange interactions \cite{ccf,cc}.
In this framework we are able to explain the existence of a pseudogap
in the optical spectra, as well as its evolution with doping and
temperature.
From the comparison between theory and experiments
we estimate a value for the local  el-ph coupling of the order 
$\lambda \approx 0.7$.

We now examine the above scenario in the light
of the theoretical DMFT results. We consider the motion of
one electron added to the half-filled antiferromagnetic
background and interacting {\em \`a la} Holstein
with local dispersionless phonons.
Due to the particle-hole symmetry on the parent Hubbard-Holstein Hamiltonian,
such problem is formally equivalent to consider one hole
in the Holstein-$t$-$J$ model. Using the linear spin-wave
approximation, we can thus write  the Hamiltonian as \cite{cc,ccf}:
\begin{eqnarray} 
H&=&\frac{t}{2\sqrt{z}}\sum_{\langle ij \rangle} 
\left(h_j^\dagger h_i a_j + {\rm h.c.}\right) 
+\frac{\tilde{J}}{2}\sum_i a_i^\dagger a_i +\nonumber \\ 
&+&g\sum_i 
h_i^\dagger h_i 
\left(b_i+b_i^\dagger\right) 
+\omega_0 \sum_i b_i^\dagger b_i,
\label{ham} 
\end{eqnarray} 
where $z$ is the coordination number,
$a^\dagger$ and $b^\dagger$ are the creation operators
for  boson spin defects and phonons respectively,
and $h^\dagger$ is the single spinless charge operator.
A local el-ph coupling constant $\lambda=g^2/\omega_0 t$ can 
be defined as the polaron energy 
in units of the half-bandwidth $t$.
The parameter $\tilde{J}$ is an effective exchange interaction.
It is linked to the microscopic $J$ by the relation
$\tilde{J}=Jm$, where
$m$ is the average on-site magnetization which, in the
$z \gg 1$ limit, is governed by the
Curie-Weiss equation $m=\tanh(\beta Jm/4)$
with N\'eel temperature $T_{\rm N}=J/4$. 
The one-particle Hamiltonian (\ref{ham})
can be thought to be representative also for the
finite doping case as far as we are in the dilute limit,
namely as far as the interparticle distance $d_{\rm P-P}$ is larger
than the spin polaron size $L_p$. For $\delta \lesssim 0.14$
this implies $L_p \lesssim 2.7$ in units of the lattice constant.
We shall discuss later the validity of this condition.
A further implication of the finite doping regime
is that the effective exchange energy $\tilde{J}$
should be considered as renormalized
by finite charge density effects, and thus doping dependent
$\tilde{J}=\tilde{J}(\delta)$.

The Holstein-$t$-$J$ Hamiltonian (\ref{ham}) is the basis
to discuss the physics of a spin/lattice polaron,
namely a travelling charge carrying along
an on-site phonon cloud and surrounded
by a local destruction of the AF background of size $L_p$.
Different properties of this object can be investigated  and,
as a general rule,
the choice of a particular theoretical approach
depends on which property is under examination and on
its feasibility to investigate it.
Quasi-particle dispersion, its
relative spectral weight and effective mass, for instance,
are intrinsically related to the coherent motion of the
polaron {\em as a whole}. On the other hand,
the internal local structure of the spin/lattice polaron, responsible for 
the optical absorption at finite frequency,
is mainly related to incoherent processes which can be safely caught
by the dynamical mean-field approach, 
%in the $z \gg 1$ limit,
where coherent motion is disregarded \cite{cc,ccf}.

One of the main intriguing characteristics of the spin/lattice
polaron that can be captured by DMFT
within the context of the Holstein-$t$-$J$ model \cite{defilippis}
is that the electron-phonon and the exchange interaction
act in a cooperative way to reduce the local hopping
amplitude, and  hence to establish polaronic
features \cite{mish1,cc,ccf,zhong,ramsak,roder,prelovsek}.
An interesting consequence of this scenario
is that lattice polaron features can be in principle switched on and off
{\em by varying the magnetic exchange coupling $\tilde{J}$ at given
el-ph coupling $\lambda$}.
Our analysis shows that this is precisely what occurs in NCCO.
On the physical ground, the effective exchange coupling
depends on the charge doping $\delta$
and on the temperature $T$, in such a way that 
increasing $\delta$ and $T$ leads to a reduction of $\tilde{J}$
and hence to an increase of the spin polaron size.
In this context we interpret the pseudogap temperature $T^*$
as the highest temperature where the spin polaron size $L_p$
is smaller than the AF correlation length $\xi_{\rm AF}$.
For $T > T^*$ ($L_p > \xi_{\rm AF}$) the charge motion does not probe
anymore the AF environment and a paramagnetic analysis with $\tilde{J}=m=0$
in enforced.
We can identify thus the experimental $T^*$ with the
N\'eel temperature $T_{\rm N}$ 
of the DMFT solution of the Holstein-$t$-$J$ model,
where the AF correlation length jumps from
$\xi_{\rm AF}=0$ for $T>T_{\rm N}$ to
$\xi_{\rm AF}=\infty$ for $T \le T_{\rm N}$
and the spin polaron size
from $L_p=\infty$ for $T>T_{\rm N}$ to a finite value for $T \le T_{\rm N}$.
Along this perspective we can determine the (doping dependent)
low temperature effective exchange energy $\tilde{J}(T=0)=J$, where $m=1$,
from the experimental temperature $T^*=T_{\rm N}=J/4$.
The estimates so obtained as function of $\delta$,
in good agreement with
experimental estimates \cite{ginsberg},
are collected in Table \ref{tab} along with other characteristic
quantities.
\begin{table}
\begin{center}
\begin{tabular}{|c|c|c|c|c|}
\hline \hline
$\delta$ & 0.05 & 0.075 & 0.1 & 0.125 \\
\hline
$T^*$ (K) & 440 & 340 & 300 & 200\\
\hline
$J$ (meV) & 152 & 117 & 103 & 69 \\
%\hline
%$J/t$ & 0.152 & 0.117 & 0.103 & 0.069 \\
\hline
$L_p$ & 0.14 & 0.19 & 0.27 & 1.01 \\
\hline
$d_{\rm P-P}$ & 4.5 & 3.7 & 3.2 & 2.8 \\
\hline \hline
\end{tabular}
\end{center}
\caption{Doping evolution of experimental data and
theoretical parameters as function of doping $\delta$.
$T^*$ is the pseudogap temperature taken from Ref. \cite{onose},
$J$ is the theoretical estimate
of low temperature exchange coupling corresponding to $T^*$,
$L_p$ is the average number of spin defects, i.e. the spin polaron size
in units of the lattice constant,
and $d_{\rm P-P}$ is the average particle-particle distance
given by $d_{\rm P-P}=1/\sqrt{\delta}$, also
in units of the lattice constant.}
\label{tab}
\end{table}
The temperature dependence of the magnetization
can thus be evaluated from the Curie-Weiss relation.
From photoemission spectroscopy we take
a half-bandwidth value $t=1$ eV for  NCCO \cite{armitage,markiewicz,ikeda},
and we set also
$\omega_0=70$ meV, consistent with the energy window
of the optical phonons in undoped cuprates.
With these values, the only free adjustable parameter
results to be the electron-phonon coupling constant $\lambda$.
We are going to show that the overall experimental scenario \cite{onose}
is naturally reproduced within the context of a spin/lattice polaron
by assuming a doping-independent el-ph coupling $\lambda=0.7$.
We would stress that this does not rule out a possible
weak dependence of $\lambda$ on doping which would
provide an even stronger agreement with the experimental data.

\begin{figure}[t] 
\begin{center} 
\includegraphics[width=7.8cm,clip=]{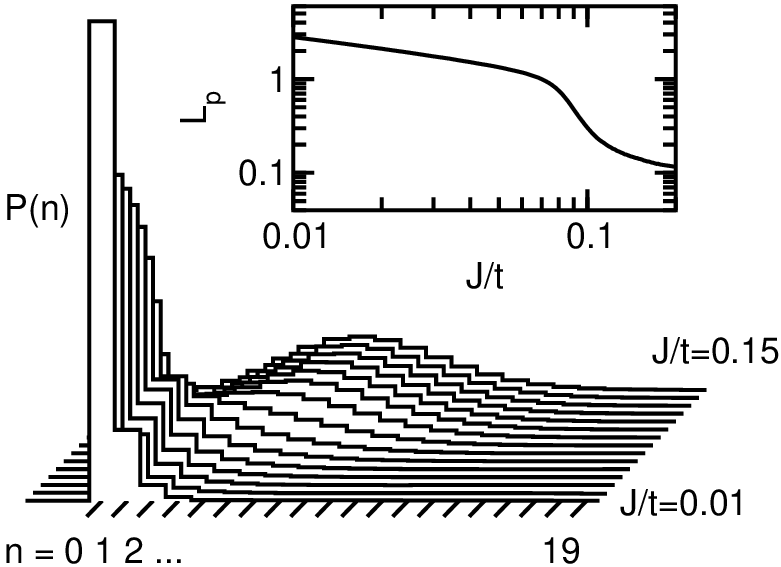} 
\end{center} 
\caption{Phonon distribution function $P(n)$ (main panel)
and spin polaron size $L_p$ (inset)
as function of exchange coupling.}
\label{f-pn} 
\end{figure}

It should be stressed that an el-ph coupling $\lambda=0.7$,
at $\omega_0/t=0.07$, is not sufficient to induce lattice polaron
effects {\em in the absence} of exchange coupling since a critical
value $\lambda_c=0.91$ is needed to establish a lattice polaron
state at $J=0$ \cite{nota1}.
This means that lattice polaron features in these compounds
can be found only in the presence of (short-range) AF correlations and they
are effectively tuned by the doping dependent exchange coupling
$J(\delta)$. In order to show and quantify this issue we plot
in Fig. \ref{f-pn} the renormalized phonon probability distribution
function $P(n)$ [$\sum_n P(n)=1$] for different values of the exchange
coupling spanning the experimental range of $J$ (see Table \ref{tab}).
The magnetic-induced lattice polaron crossover occurs for
$J/t \simeq 0.083$ ($J=83$ meV in energy units), where the $P(n)$
changes abruptly from a distribution peaked  at $n=0$ to
a multiphonon broad structure at finite $n$, typical of a
polaronic lattice distortion.
The strong interplay between lattice and magnetic degrees of
freedom is also witnessed by the dependence of the
spin polaron size $L_p$  on $J$ at $T=0$, obtained by the
Hellmann-Feynman theorem\cite{cc}, as shown
in the inset of Fig. \ref{f-pn}. Here the onset of
the lattice polaron for $J/t \gtrsim 0.083$ leads to a
further self-trapping localization of the charge and to
an abrupt drop of $L_p$ in the polaronic regime.
Comparing the typical values of the exchange energy estimated
from the experimental data reported in Table \ref{tab},
we can predict lattice polaron features, accompanied
by a very small amount of spin defects, for
dopings $\delta = 0.05, 0.075,0.1$ while at the larger doping
$\delta=0.125$ spin/lattice polaron features are
already remarkably reduced.
It is also comforting to check that for all the dopings considered
in the present analysis $\delta = 0.05-0.125$,
both on the polaronic and on the non-polaronic side,
the estimate of the spin polaron size $L_p$
is always smaller than the average particle-particle distance $d_{\rm P-P}$
(see Table \ref{tab}),
confirming the validity of the present one-particle analysis
in the dilute limit.

\begin{figure}[t] 
\begin{center} 
\includegraphics[width=7.8cm,clip=]{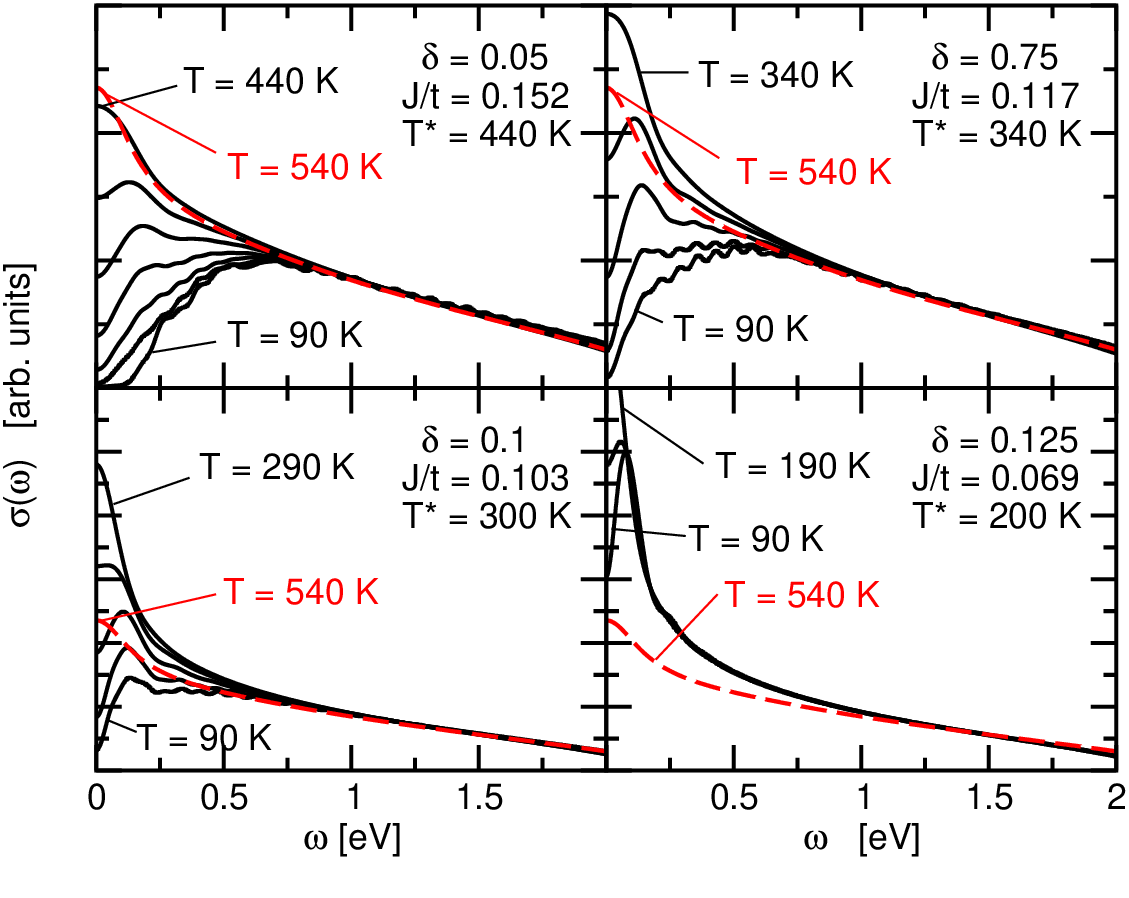} 
\end{center} 
\caption{(color online) Temperature evolution of the optical conductivity
$\sigma(\omega)$ for different exchange couplings (dopings) and for
($T \le T^*$):
$T =  90, 140, 190, 240, 290, 340, 390, 440$ K). 
%Upper lines correspond to higher temperatures.
Also shown is the optical conductivity at $T=540$ K in the normal state
(dashed red line).}
\label{f-oc} 
\end{figure}

Let us now discuss the effects of the magnetic induced
lattice polaron formation on the optical conductivity.
In Fig. \ref{f-oc} we plot $\sigma(\omega)$ evaluated
for the different exchange couplings corresponding to different doping
levels (see Table \ref{tab}).
We should stress, once more, that, due to the local approximation enforced
by the $z=\infty$ DMFT limit, only incoherent contributions to the OC
are here taken into account, so that no Drude-like peak is present.
Disregarding the coherent Drude peak, the resemblance
with the experimental data of Ref. \cite{onose} is remarkable.
In particular we observe at low temperature a 
broad incoherent shoulder in the MIR energy range which
characterizes the pseudogap regime.
Such feature represents the natural counterpart, in the
optical data, of the broad multiphonon structure
responsible for the incoherent peak observed
in ARPES, and is a natural signature of the
onset of the lattice polaron. Consistent with this picture,
the energy scale of such broad MIR band is reduced 
as polaronic features are weakened upon doping (Fig. \ref{f-oc}),
which causes a closing of the optical pseudogap.
%For the largest doping $\delta=0.125$ a polaronic
%peak is only visible at very low energy $\omega \approx 0.08$ eV.
Also interesting is the role
of  temperature, also reported in Fig. \ref{f-oc}.
While the magnitude of the exchange coupling $J$ rules
the energy scale of the MIR band \cite{ccf} at low temperature,
and hence drives the {\em closing} of the pseudogap with $\delta$,
increasing the temperature for $T < T^*$ 
leads to a {\em filling} of the pseudogap
itself (see in particular the upper panels in Fig. \ref{f-oc}), 
in good agreement with Ref. \cite{onose}.
Eventually for $T> T^*$ there is a considerable smoothing of the
OC features with respect to the spectra at $T \approx T^*$.

\begin{figure}[t] 
\begin{center} 
\includegraphics[width=7.8cm,clip=]{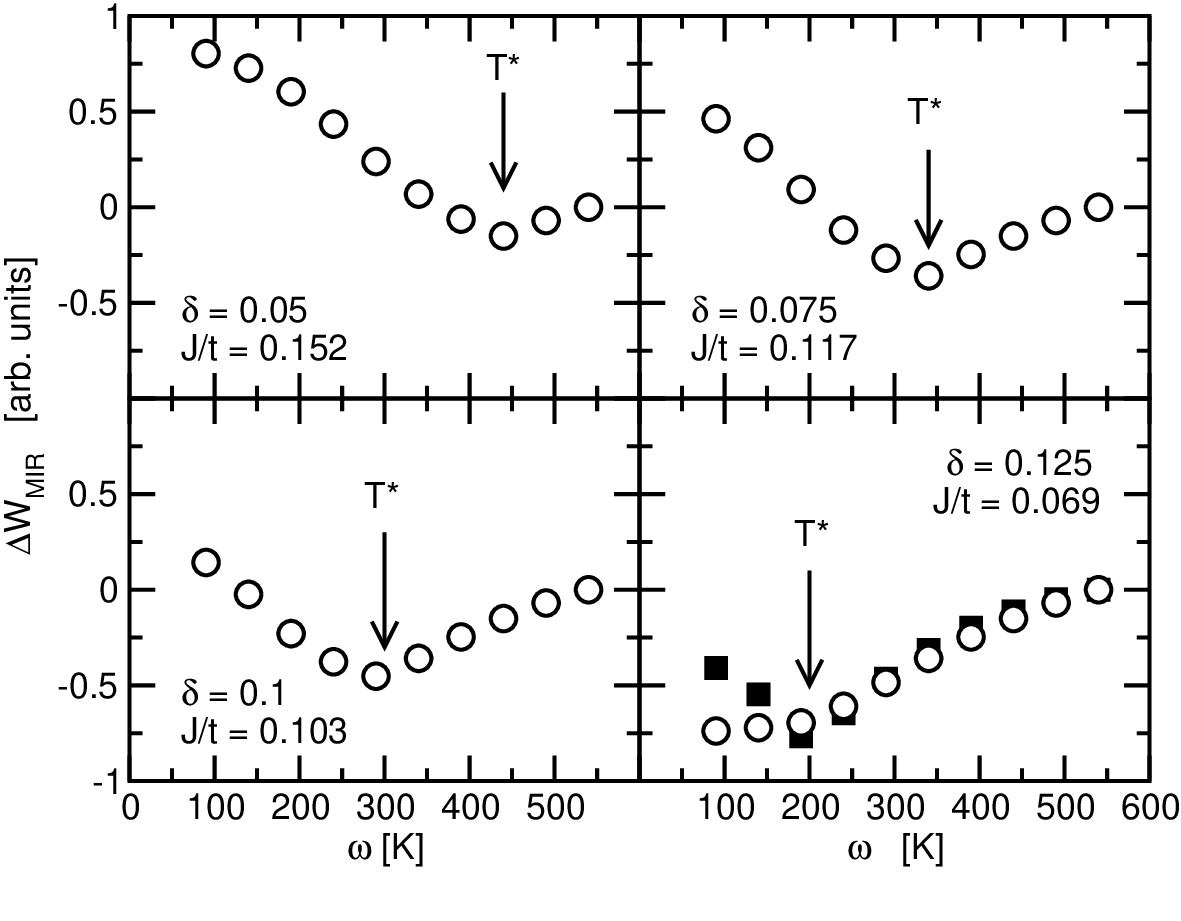} 
\end{center} 
\caption{Temperature dependence of $\Delta W_{\rm MIR}$
for different exchange coupling corresponding to
different doping levels. Empty circles and black squares
are obtained by integrating $\sigma(\omega)$
in the [0.12,1] eV and in the [0,1] eV energy window,
respectively.}
\label{f-w} 
\end{figure}

Such temperature evolution can be nicely traced by evaluating the
``MIR optical spectral weight'' $W_{\rm MIR}$, which was defined
in Ref. \cite{onose} as the optical spectral weight between a low energy
cut-off at $\omega_{\rm low}=0.12$ eV and a high-energy isosbestic 
point. Quite notably, the temperature evolution of such
spectral weight does not show any signature of the AF long-range order
at the N\'eel temperature $T_{\rm N}^{\rm exp}$, whereas a non-monotonic kink
at a higher temperature $T^*$ is clearly visible. 
This observation can be naturally
explained within the present (local) polaronic approach, 
since the {\em local}
polaron processes responsible for the incoherent shoulder
in Fig. \ref{f-oc} cannot distinguish between a short-range and long-range
AF ordering as long as $L_p < \xi_{\rm AF}$.
For the same reasons,  anomalies in $W_{MIR}$ could be expected
at $T \gtrsim T^*$ ($T \gtrsim T^{\rm N}$ in our analysis)
where $L_p \gtrsim \xi_{\rm AF}$ and the concept
itself of a spin polaron breaks down.
In order to quantify this issue,
in the absence of a corresponding isosbestic point in our data,
we define  $W_{\rm MIR}$ as the optical spectral weight between
$\omega_{\rm low}=0.12$ eV and a  fixed high-energy 
cut-off at $\omega_{\rm high}=1$ eV. We have checked that different choices
of $\omega_{\rm high}$ do not affect qualitatively our results
for $\delta=0.05, 0.075, 0.1$. The dependence on
$\omega_{\rm low}$ and $\omega_{\rm high}$ for $\delta=0.125$
is explicitly discussed later.
The temperature evolution of $W_{\rm MIR}$ is shown in Fig. \ref{f-w},
where we plot $\Delta W_{\rm MIR}=
W_{\rm MIR}(T=540\mbox{ K})-W_{\rm MIR}(T)$
to have a direct comparison with Ref. \cite{onose}.
The resemblance with the experimental data is striking.
For all dopings we find that the reduction of
$\Delta W_{\rm MIR}$ as function of $T$, which reflects the filling
of the pseudogap, is accompanied by a 
change of behavior at $T^*$. 
%and by
%a subsequent {\em increase} of $\Delta W_{\rm MIR}$. 
We associate such different behaviors to two different regimes.
In the high temperature regime $T>T^*$ polaronic effects
are not operative: increasing the temperature leads to the
usual smearing of the overall shape of the optical spectra,
and thus to a depletion of the spectral weight in the MIR energy window
(i.e. an increase of $\Delta W_{\rm MIR}$). In the
polaronic regime at $T<T^*$ this effect is counteracted
by the stronger temperature dependence of the polaronic effects.
The latter leads to a filling of the pseudogap, and therefore to
an increase of $W_{\rm MIR}$ (a decrease of $\Delta W_{\rm MIR}$)
upon increasing the temperature.
The competition between these two effects gives rise to a kink
in the temperature evolution of $\Delta W_{\rm MIR}$.
Such kink is not clearly visible at the highest doping
$\delta=0.125$ in Fig. \ref{f-w}. This is  due
to the fact that, in contrast to
Ref. \cite{onose},  in our data at $\delta=0.125$ the weak polaronic MIR peak
is at $\omega \approx 0.08$ eV, out  of the range [0.12:1] eV
of the spectral weight integration.
The agreement with Ref. \cite{onose} 
is recovered by extending the integration of 
$\Delta W_{\rm MIR}$ down to  $\omega_{\rm low}=0$ (black squares
in the last panel of
Fig. \ref{f-w}). 
%also the $\Delta W_{MIR}$ obtained
%by integration between $\omega_{\rm low}=0$ and $\omega_{\rm low}=0.1$ eV
%(black squares), showing also in this case the kink
%observed in Ref. \cite{onose}.

In conclusion, in this paper we have analyzed the optical data
of the electron-doped cuprate 
Nd$_{2-x}$Ce$_x$CuO$_4$ as function of doping and temperature
by means of the DMFT results for one electron in the Holstein-$t$-$J$ model.
We have shown that the origin of the MIR band, the optical pseudogap 
as well as its doping and temperature dependence can be naturally explained in 
terms of a polaronic scenario
resulting from the joint cooperation of the el-ph and exchange interactions.
From our analysis we estimate an el-ph coupling
of the order of $\lambda \simeq 0.7$. Our results illustrate the
relevance of the positive interplay between lattice and spin degrees
of freedom in electron-doped cuprates, where the  AF ordering 
extends over a large doping window, making these compounds ideal systems where
to investigate such effects.

We thank Y. Onose, Y. Tokura,
A.S. Mishchenko, V. Cataudella and G. De Filippis
for useful discussions. We also thank
Y. Onose and Y. Tokura for having made available
their experimental data.
E.C. and S.C. acknowledge also financial support from the
Research Program MIUR-PRIN 2005.


\begin{thebibliography}{99}
\bibitem{nota1}
We define in an unambiguous way
a lattice polaron instability, for generic $\lambda$ and $J$,
by the condition $P(n=0)\le P(n \neq 0)$, where
$P(n)$ is the phonon distribution function
(see Ref. \cite{cc}).
\end{thebibliography}
\end{document}